\documentclass[doublecol]{epl2}

\title{A Krein Quantization Approach to Klein Paradox}
\shorttitle{A Krein Quantization Approach to Klein Paradox} 

\author{Farrin Payandeh\inst{1}\thanks{e-mail: f$\_$payandeh@pnu.ac.ir} \and Toradj Mohammad Pur\inst{2} ,  Mohsen Fathi\inst{1}\thanks{e-mail: mohsen.fathi@gmail.com}
\and Zahra Gh. Moghaddam\inst{3}\thanks{e-mail:
z.gh.moghaddam@gmail.com}} \shortauthor{F. Payandeh \etal}

\institute{
  \inst{1} Department of Physics, Payame
Noor University, PO BOX 19395-3697 Tehran, Iran\\
  \inst{2} Department of Physics, Payame Noor University,
Tabriz, Iran\\
   \inst{3} Department of Physics, Islamic
Azad University, Central Tehran Branch, Tehran, Iran

 }

\abstract{ In this paper we first introduce the famous Klein
paradox. Afterwards by proposing the Krein quantization approach
and taking the negative modes into account, we will show that the
expected and exact current densities, could be achieved without
confronting any paradox.}

\begin{document}

\maketitle

\section{Introduction and Motivation}
A famous problem in quantum mechanics concerns with particles
confined to a barrier. This is where the famous tunnelling effect
arises. For relativistic quantum particles behind such potential
barriers, one can use the Klein-Gordon equation
\begin{equation}\label{KG}
\Big[ ih\frac{\partial}{\partial t}-V\Big]^2\varphi=-\nabla^2
c^2\varphi+m_0^2c^4\varphi,
\end{equation}
which describes a plane wave solution for the relativistic
particles, appropriating them a total energy $E$, before and after
the barrier. Also the potential $V$ is usually supposed to be of
form of a step function. Here an important notion which is
critical in quantum mechanical calculations, would be the
conservation of probability current or charge current \cite{1}.
Also no particle flux in the barrier is supposed to be existed in
the positive direction (Region B in Figure 1).

Let us begin with the Klein-Gordon equation. The total Energy from
special relativity turns out to be
\begin{equation}\label{E-1}
E^2=p^2c^2+(m_0c^2)^2,
\end{equation}
where $p$ is the linear momentum. If any potential was available,
this energy could be interpreted as
\begin{equation}\label{E-2}
E+V\rightarrow ih\frac{\partial}{\partial t}.
\end{equation}

\begin{figure}[htp]\label{F1}
\center{\includegraphics[width=8cm]{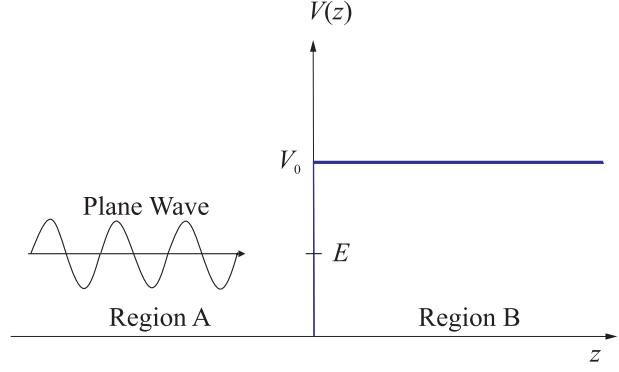} \caption{\small{The
relativistic quantum particles, described by a plane wave, before
a potential barrier. The two regions A and B are also notated.}}}
\end{figure}
Equation (\ref{KG}) possesses regular solutions for the field
$\varphi$ in Region A,
\begin{equation}\label{phiA}
\varphi_{\textmd A}=e^{-\frac{i}{h}(Et-p.z)}+R
e^{-\frac{i}{h}(Et+p.z)},
\end{equation}
and Region B:
\begin{equation}\label{phiB}
\varphi_{\textmd B}=Te^{\frac{i}{h}(Et-p^\prime.z)},
\end{equation}
where $R$ and $T$ are two constants, related to the amplitude of
the wave, and have to be identified.

\subsection{Klein Paradox from Klein-Gordon equation}
From equation (\ref{KG}), and also regarding the definitions for
energy in (\ref{E-1}) and (\ref{E-2}), the total linear momentum
can be written as
\begin{equation}\label{p-1}
p=\sqrt{E^2-m^2c^4}
\end{equation}
and
\begin{equation}\label{p-2}
p^\prime=\pm\sqrt{(E-V)^2-m^2c^4}.
\end{equation}
This could help us to categorize the total linear momentum with
respect to the relations between the the total energy and the
potential. The following cases arise:

\begin{itemize}
\item{for weak potentials, where $V<E-mc^2$, the momentum
$p^\prime$ would be real and only opts positive values.}

\item{for an intermediate potential, $E-mc^2<V<E+mc^2$, $p^\prime$
takes imaginary values, providing unstable waves.}

\item{when the potential is strong, i.e. $V>E+mc^2$, then $p'$ is
real and exhibits non-classical behaviors.}

\end{itemize}
Now a question arises: \\

{"\it{How can we guarantee the charge current conservation?}"}\\

To deal with this question, we initially have to determine the
values for $R$ and $T$ in (\ref{phiA}) and (\ref{phiB}). To do
this, we are expected to apply the continuity condition for
$\varphi$ and its derivative at $z=0$. In other words we set
$$\varphi_A|_{z=0}=\varphi_B|_{z=0},$$
$$\varphi^\prime_A|_{z=0}=\varphi^\prime_B|_{z=0}.$$
Therefore we get a system of linear equations.
\begin{equation}\label{R,T-1}
1+R=T,$$$$ p(1-R)=Tp^\prime.
\end{equation}
Solving the system in (\ref{R,T-1}), gives the following values
for $R$ and $T$:
\begin{equation}\label{R,T-2}
R=\frac{p-p^\prime}{p+p^\prime},$$$$ T=\frac{2p}{p+p^\prime}.
\end{equation}
Now we get back to the conservation of charge current. The charge
current of a massive scalar field is defined by
\begin{equation}\label{j-1}
\overrightarrow{j}=\frac{1}{2im}\Big(\varphi^*\overrightarrow{\nabla}\varphi-\varphi\overrightarrow{\nabla}\varphi^*\Big),
\end{equation}
from which the current for the fields in (\ref{phiA}) and
(\ref{phiB}) are derived as:
\begin{equation}\label{jA}
 {j_A} = \frac{h}{{2im}}\Big({e^{\frac{i}{h}(Et - p.z)}} + {{{R}}^{{*}}}{}{e^{\frac{i}{h}(Et + p.z)}}\Big)\Big(\frac{{ip}}{h}{e^{\frac{{ - i}}{h}(Et - p.z)}} - {{R }}\frac{{ip}}{h}{e^{\frac{{ - i}}{h}(Et + p.z)}}\Big)
 $$$$
   - \frac{h}{{2im}}\Big({e^{\frac{{ - i}}{h}(Et - p.z)}} + {{R }}{e^{\frac{{ - i}}{h}(Et + p.z)}}\Big)\frac{{ip}}{h}\Big( - {e^{\frac{i}{h}(Et - p.z)}} + {{{R}}^{{*}}}{}{e^{\frac{i}{h}(Et +
   p.z)}}\Big),$$$$
   < {j_A} >  =\frac{p}{m}(1 - {\left| R
   \right|^2}),
\end{equation}
and
\begin{equation}\label{jB}
 {j_B} = \frac{h}{{2im}}\Big[{T^*}{e^{\frac{i}{h}(Et - p^{\prime *}.z)}}\frac{{ip^\prime }}{h}T{e^{\frac{{ - i}}{h}(Et - p^\prime .z)}}
 - T{e^{\frac{{ - i}}{h}(Et - p^\prime .z)}}( - \frac{{ip^\prime }}{h}){T^*}{e^{\frac{i}{h}(Et - p^{\prime
 *}.z)}}\Big],$$$$
\left\{ \begin{array}{l}
 < {j_B} >  = \frac{{p\prime }}{m}\left| {{T}} \right|^2,{\rm{    }}\,\,\,\,\,\,\,\,\,\,p^\prime \,\,{\rm{
 is\,\,
 real}}\\\\
 < {j_B} >  = 0,\,\,\,\,\,\,\,\,\,\,{\rm{             }}p^\prime \,\,{\rm{ is\,\, imaginary}}
\end{array} \right..
\end{equation}
Also one can define an average incident current, due to the linear
momentum and mass of the field as
\begin{equation}\label{j-in}
<j_{inc}>=\frac{p}{m}.
\end{equation}
This will help us to investigate the ratios between $p$ and
$p^\prime$, as functions of the reflection coefficient $R$, and
the transmission coefficient $T$. Form (\ref{R,T-2}) and the
definitions in (\ref{jA}) and (\ref{jB}) we have:
\begin{equation}\label{j-j-in}
|R|^2=\frac{|j_A-j_{inc}|}{j_{inc}}=\frac{|j_R|}{j_{inc}},$$$$
|T|^2=\frac{p}{p^\prime}\frac{j_B}{j_{inc}}.
\end{equation}
And it is always expected that $R+T=1$. Note that, for an
intermediate potential, $R=1, \,\,T=0$ and for a strong potential
\cite{125, 3}
\begin{equation}\label{R,T-strong}
R=\Big(\frac{p+|p^\prime|}{p-|p^\prime|}\Big)^2,$$$$
T=-\frac{4p|p^\prime|}{(p-|p^\prime|)^2}.
\end{equation}
One can observe that, for both cases the condition $R+T=1$ is
satisfied. However, equation (\ref{R,T-strong}) asserts that $R>1$
and $T<0$; that is the reflected current is bigger than incident
current, or the transmitted current is opposite in charge to
incident current. This is what we know as the Klein paradox.

Historically, this result was obtained by Oskar Klein in 1929
\cite{2} and since then, much effort has been devoted to this
problem by stating that this unexpected reflected current is
because of some extra particles which are being supplied by the
potential, or this negative transmitted current is caused by
another type of particles, possessing opposite charges \cite{3,4}.
This explanation of Klein paradox was based on Klein-Gordon
equation. Now let us have another approach through the Dirac
equation.

\subsection{Klein Paradox from Dirac equation}
According Figure 1, one can consider two operating equations
\cite{5}. One for Region A ($z<0$):
\begin{equation}\label{z<0}
(c\alpha\hat p+\beta mc^2)\psi=E\psi,
\end{equation}
and one for Region B ($z>0$):
\begin{equation}\label{z>0}
(c\alpha\hat p+\beta mc^2)\psi=(E-V_0)\psi.
\end{equation}
Also it would be possible to write down an incident wave function
in region A.
\begin{equation}\label{inc-wave}
\psi_{inc}=\alpha\left(\begin{array}{l}
\,\,\,\,\,\,\,1\\\,\,\,\,\,\,\,0\\\frac{pc}{E+mc^2}\\\,\,\,\,\,\,\,0\end{array}\right)e^{\frac{ipz}{\hbar}},
\end{equation}
from which the reflected and the transmitted spinors are derived
as
\begin{equation}\label{ref-spin}
\psi_R=b\left(\begin{array}{l}
\,\,\,\,\,\,\,1\\\,\,\,\,\,\,\,0\\\frac{-pc}{E+mc^2}\\\,\,\,\,\,\,\,0\end{array}\right)e^{\frac{-ipz}{\hbar}}+b^\prime\left(\begin{array}{l}
\,\,\,\,\,\,\,0\\\,\,\,\,\,\,\,1\\\,\,\,\,\,\,\,0\\\frac{pc}{E+mc^2}\end{array}\right)e^{\frac{-ipz}{\hbar}},
\end{equation}
\begin{equation}\label{tran-spin}
\psi_T=d\left(\begin{array}{l}
\,\,\,\,\,\,\,\,\,\,\,\,1\\\,\,\,\,\,\,\,\,\,\,\,\,0\\\frac{p^\prime
c}{E-V_0+mc^2}\\\end{array}\right)e^{\frac{ip^\prime
z}{\hbar}}+d^\prime\left(\begin{array}{l}
\,\,\,\,\,\,\,\,\,\,\,\,0\\\,\,\,\,\,\,\,\,\,\,\,\,0\\\frac{-p^\prime
c}{E-V_0+mc^2}\end{array}\right)e^{\frac{ip^\prime z}{\hbar}}.
\end{equation}
Note that,
$$p^\prime c=\sqrt{(V_0-E)^2-m^2c^4}.$$
Now consider the case of existence of a strong potential,
$V_0>E+mc^2$. Same as in the previous subsection, applying the
continuity condition on spinors at the boundary
$\psi_{inc}+\psi_R=\psi_T$, we get \cite{6}:
\begin{equation}\label{B.C.inc+R=T}
a+b=d,$$
$$b^\prime=d^\prime,$$
$$\frac{pc}{E+mc^2}a-\frac{pc}{E+mc^2}b=\frac{-p^\prime
c}{V_0-E-mc^2}d,$$
$$\frac{pc}{E+mc^2}b^\prime=\frac{p^\prime c}{V_0-E-mc^2}d^\prime.
\end{equation}
Note that, the case $b^\prime=d^\prime=0$, means no spin flip.

In order to get back to the Klein paradox, let us write down the
probability currents. The probability current can be written as
\begin{equation}\label{j(x)}
j(x)=c\psi^\dagger(x)\alpha_3\psi(x),
\end{equation}
from which, using (\ref{B.C.inc+R=T}), we can derive the
incidental, reflective and transmitted currents.
\begin{equation}\label{j(x),inc,R,T}
j_{inc}=aa^*\frac{2pc^2}{E+mc^2},$$
$$j_R=-bb^*\frac{2pc^2}{E+mc^2},$$
$$j_T=-dd^*\frac{2p^\prime c^2}{V_0-E-mc^2}.
\end{equation}
This could help us to identify the reflection and transmission
coefficients \cite{3,7}.
\begin{equation}\label{R,T-Dirac}
R=\frac{j_R}{j_{inc}}=-\frac{bb^*}{aa^*}=\frac{(1+r)^2}{(1-r)^2},$$
$$T=\frac{j_T}{j_{inc}}=-\frac{4r}{(1-r)^2}.
\end{equation}
Dealing with equation (\ref{R,T-Dirac}), one can see that for
$r=\sqrt{\frac{(V_0-E+mc^2)(E+mc^2)}{(V_0-E-mc^2)(E-mc^2)}}$ we
have $r\geq1$. Once again, $R>1,\,\,\,T<0$; the reflected current
is greater than the incident current. This means that we have been
confronted the Klein paradox.

In this paper we concern about the mentioned paradox, however from
another viewpoint, namely from Krein space quantization, which
will be introduced in the next section. An important idea to
surmount Klein paradox is that it is supposed that the potential
energy increases the negative energy of the electron, to a
positive energy state, creating a positive hole (positron) behind
it. The hole is attracted towards the potential while the electron
is repelled far from it. This process is stimulated by the
incoming electron (see Figure 2). However, in this article we
suggest that we could keep the negative energies as viable
energies. Through Krein quantization approach, we put that these
negative energies were possible to be included in our calculations
and just like what was asserted, they essentially could be
regarded as un-physical particles and antiparticles. Having them,
the energy conservation is also guaranteed. First of all, let us
have an overview on Krein quantization.
\begin{figure}[htp]\label{F2}
\center{\includegraphics[width=10cm]{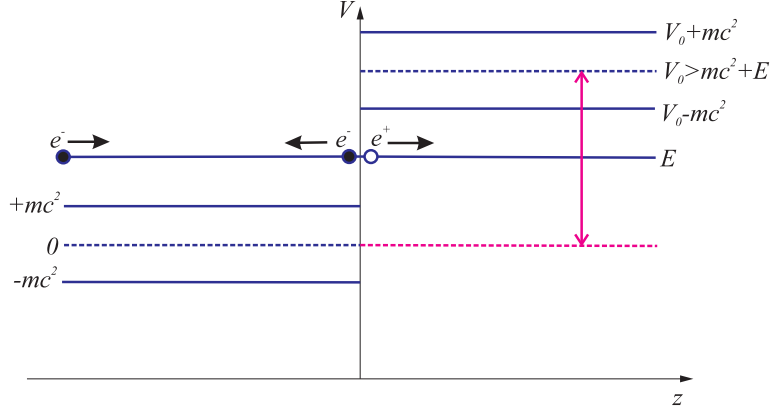} \caption{\small{The
incoming negative electron can be regarded as a reflected negative
electron with same energy, and a transmitted positron.}}}
\end{figure}

\section{Krein Quantization}
As it was discussed in the previous section, there would be some
unexpected negative energies in transmitted fermions trough the
barrier. Initially, Dirac proposed to keep these negative states.
After that, much efforts have been devoted to construct a viable
theory for appropriate interpretations of negative energies. What
we are concerning about here, is a new method, called Krein
quantum field theory, being able to use the so-called negative
energies. Krein quantization is based on removing the divergences,
caused by singularities in Green's function \cite{8}.

Let us begin with a scalar field quantization in the form
\cite{9,10,127,11}
\begin{equation}\label{Kr-1}
\phi(x)=\frac{1}{\sqrt{2}}\Big[\phi_p(x)+\phi_n(x)\Big],
\end{equation}
where
$$\phi_p(x)=\int d^3\overrightarrow{k}\,\,\Big[     a(\overrightarrow{k})u_p(\overrightarrow{k},x)+a^\dagger(\overrightarrow{k})u^*_p(\overrightarrow{k},x)       \Big],$$
and
$$\phi_n(x)=\int d^3\overrightarrow{k}\,\,\Big[     b(\overrightarrow{k})u_n(\overrightarrow{k},x)+b^\dagger(\overrightarrow{k})u^*_n(\overrightarrow{k},x)       \Big].$$
Here, the lower the indices $p$ and $n$, respectively referring to
the positive and negative states (or modes). The positive mode is
the usual scalar field and the negative one, which here we are
about to consider, would be the regularization field. As we
mentioned above, the divergences in quantum field theory are
caused by the Green's function singularities. This Green's
function is defined as a time-ordered product \cite{133,134,135}.
\begin{equation}\label{Kr-2}
iG_T(\overrightarrow{x},\overrightarrow{x}')=<0|T\phi(x)\phi(x^\prime)|0>=\textmd{Re}[G_F(\overrightarrow{x},\overrightarrow{x}')],
\end{equation}
where $G_F(\overrightarrow{x},\overrightarrow{x}')$ is the Feynman
Green function \cite{11}. According to this, the time-ordered
product propagator in the Feynman gauge for the vector field in
Krein space is given by \cite{11,139}:
\begin{equation}\label{propagator-1}
<D_{\mu\nu}^T(x,x')>=-\eta_{\mu\nu}<G_T(x,x')>.
\end{equation}
The most essential notion of Krein quantization would be its
impact on the solutions of Dirac equation. The Dirac field in
Krein space is written in the following form (for a detailed
discussion see \cite{8}):
\begin{equation}\label{Dirac-Kr-1}
\Phi_{\textmd{D-K}}(\overrightarrow{x})=\frac{1}{\sqrt{2}}\int
d^3\overrightarrow{k}\sum_{i=1,2}\Big[(b_{\overrightarrow{k}i}+c^\dagger_{\overrightarrow{k}i})\mathbf{P}^i(\overrightarrow{k},\overrightarrow{x})
+(d^\dagger_{\overrightarrow{k}i}+a_{\overrightarrow{k}i})\mathbf{N}^i(\overrightarrow{k},\overrightarrow{x})\Big],
\end{equation}
in which the the modes $\mathbf{P}^i$ and $\mathbf{N}^i$ are
defined as
\begin{equation}\label{Pi}
\mathbf{P}^i(\overrightarrow{k},\overrightarrow{x})=\sqrt{
\frac{m}{(2\pi)^3\omega_{\overrightarrow{k}}}
}\,\,\mathbf{p}^i(\overrightarrow{k})e^{-i\overrightarrow{k}.\overrightarrow{x}}
\end{equation}
for positive energies, and
\begin{equation}\label{Ni}
\mathbf{N}^i(\overrightarrow{k},\overrightarrow{x})=\sqrt{
\frac{m}{(2\pi)^3\omega_{\overrightarrow{k}}}
}\,\,\mathbf{n}^i(\overrightarrow{k})e^{i\overrightarrow{k}.\overrightarrow{x}}
\end{equation}
for negative energies. Here it is necessary to indicate the
notions of the operators in (\ref{Dirac-Kr-1}). We introduce
\cite{136}
\begin{itemize}
    \item $b$: is the annihilation operator of one-particle (or
    one-antiparticle) state with positive energy.

    \item $c^\dagger$: is the creation operator of one-particle (or
    one-antiparticle) state with positive energy.

    \item $d^\dagger$: is the creation operator of one-particle (or
    one-antiparticle) state with negative energy.

    \item $a$: is the annihilation operator of one-particle (or
    one-antiparticle) state with negative energy.
\end{itemize}
Also the time-ordered propagator is defined as
\begin{equation}\label{ST}
S_T(\overrightarrow{x},\overrightarrow{x}')=(i\not\partial+m)G_T(\overrightarrow{x},\overrightarrow{x}'),
\end{equation}
in which the Green function
$G_T(\overrightarrow{x},\overrightarrow{x}')$ has been presented
in (\ref{Kr-2}). These two modes would be the key point in our
approach to discuss the Klein paradox, which we will deal with in
the next section. This suggestion is based on this belief that,
although have not been correlated to physical concepts, the
negative norm states are still appearing in the mathematical
procedures, together with the positive energies; as we will see in
the next section, they have an important impact on the results.
Therefore, the un-physical (or virtual) particles, may appear to
have physical meanings in the future, however, we are dealing with
the mathematical results and according to Feynman's phrase, we are
not "{\textit{hiding the rushes under the
carpet}}". \\

Through Krein quantization, we are asserting that solutions are
corresponding to the particles and antiparticles of positive
energies (physical particles of positive states) and those of
negative energies (un-physical particles of negative stats).
Therefore, in our approach, we shall maintain all 4 solutions, in
order to having all physical and un-physical particles and
antiparticles. \cite{9,10,133,132}.\\

As stated by Dirac, "\textit{negative energies and probabilities
should not be considered as nonsense. They are well-defined
concepts mathematically, like a negative sum of money, since the
equations which express the important properties of energies and
probabilities, can still be used when they are negative. Thus,
negative energies and probabilities should be considered simply as
things which do not appear in experimental results. The physical
interpretation of relativistic theory involves these things and is
thus in contradiction with experiment. We therefore, have to
consider ways of modifying or supplementing this interpretation}."
\cite{123,124}.

\section{Explaining Klein Paradox through Krein Quantization}
Maintaining an overlook on Klein paradox via Klein-Gordon and
Dirac solutions, now let us present a recently proposed
explanation for this paradox, which is based on Krein
quantization. In section one, it was asserted that among the
electrons of positive energies, which are coming down onto the
potential barrier, some are being reflected from the barrier
(travelling along $-z$ direction), and some could be assumed to be
the passing positrons travelling along $+z$ direction (see Figure
2). However, a crucial point has to be the reflected electrons of
negative energy, and this is what we are about to consider in our
new approach. According to Dirac's equation, there are four sets
of solutions available, including up and down spin electrons of
positive energy, and up and down spin electrons of negative
energy. In all cases, which we have dealt with, the incidental
electron current (Region A in Figure 1) always possesses positive
energies (having either up or down spins). This means that the
so-called negative energies have been ruled out.

The technical point in Krein quantization approach, is keeping all
the sets of solutions, even for negative energies. Let us see what
this procedure provides us. Concerning with the electrons with
negative energies, one can discover that the reflected and
transmitted electrons are covering all the incidental current, and
this means that the transmitted positrons (or transmitted
positrons of positive energy) are no longer necessary to be
considered to surmount the problem of negative current density.
Strictly speaking, what we are going to put here, is that problems
like backwardly moving electrons of negative current which are
leading to Klein paradox, are indeed arising from the fact that we
have ignored the electrons of negative energies. This lack of
initial data, will command us to consider some reflected positrons
and transmitted electrons of positive energy. However if we had
maintained all the Dirac's solutions, we could have protected our
selves of this explanation, since this leads to an equality
between the incidental and the total reflected currents, for
either of the electrons and the positrons (see Figure 3). This
means that, we could overcome the Klein paradox. In other words,
through this approach, no paradox will remain to be explained.
\begin{figure}[t] \label{F3}
\center{\includegraphics[width=11cm]{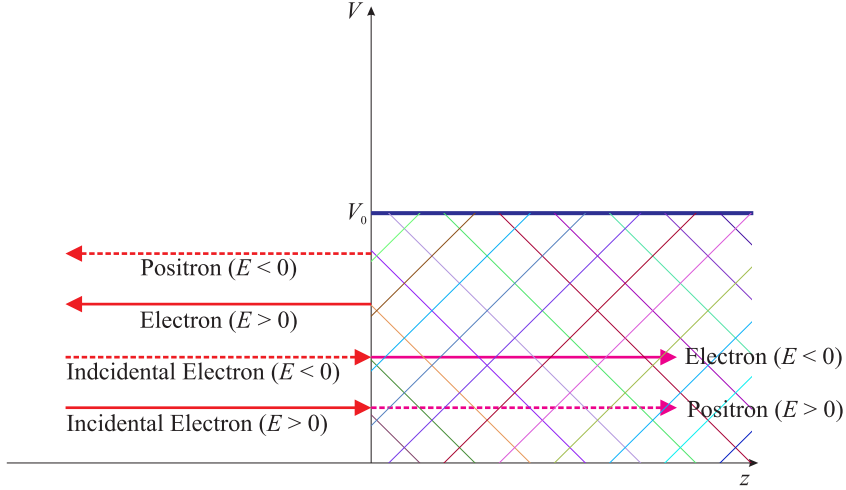} \caption{\small{The
incidental electrons or positrons having positive and negative
energies.}}}
\end{figure}
To prove our recent statements, recall the parameter $r$ related
to the reflected current from equation (\ref{R,T-Dirac}). As it
was mentioned in section one, we have
\begin{equation}\label{r}
r=\sqrt{\frac{(E-V_0-mc^2)(E+mc^2)}{(E-V_0+mc^2)(E-mc^2)}}
\end{equation}
for which we have $r>1$. Letting the negative modes to contribute
in our calculations, we define another parameter, namely
$r^\prime$, to retain the correlations between the so-called
negative states and the reflected current. If $E<0$, then the
negative states choose
\begin{equation}\label{r'}
r'=\sqrt{\frac{(-E-V_0-mc^2)(-E+mc^2)}{(-E-V_0+mc^2)(-E-mc^2)}},
\end{equation}
as the reflection parameter for electrons of negative energies.
Note that
$$r r'=k,\,\,\,\,\,\,\,\,\,\, k>1.$$
Having this, we can investigate the ratios between the incidental,
reflected and transmitted currents; i.e.
$\frac{|j'_{T}|}{|j'_{inc}|}$ and $\frac{|j'_R|}{|j'_{inc}|}$.
Equation (\ref{r'}) yields
$$-E-V_0+mc^2<0\,\,\,\,\,\,\,\,\,\,\,\,\,\textmd{or}\,\,\,\,\,\,\,\,\,\,\,\,-E<V_0-mc^2.$$
Multiplying $r$ and $r'$ from (\ref{r}) and (\ref{r'}), we get
\begin{equation}\label{rr'}
rr'=\sqrt{\frac{(E-V_0-mc^2)(E+V_0+mc^2)}{(E-V_0+mc^2)(E+V_0-mc^2)}}.
\end{equation}
Simplifications give
\begin{equation}\label{k-1}
k=\sqrt{   \frac{E^2-(V_0+mc^2)^2}{E^2-(V_0-mc^2)^2}  }.
\end{equation}
It turns out that always $k>1$ and for $V_0\rightarrow\infty$,
$k\rightarrow 1$. Now having $r'$ and $k$, we can derive same
relations for density currents as they are in (\ref{R,T-Dirac}),
for the negative modes.
\begin{equation}\label{R',T'-Dirac}
R'=\frac{|j'_R|}{|j'_{inc}|}=\frac{(1+r')^2}{(1-r')^2}=\frac{\Big(1+\frac{k}{r}\Big)^2}{\Big(1-\frac{k}{r}\Big)^2}=\frac{(k+r)^2}{(k-r)^2},$$$$
T'=\frac{|j'_{T}|}{|j'_{inc}|}=\frac{4r'}{(1-r')^2}=\frac{4\Big(\frac{k}{r}\Big)}{\Big(1-\frac{k}{r}\Big)^2}=\frac{4kr}{(k-r)^2}.
\end{equation}
Now to advocate our previously claim, that having all the modes
could give us equal numbers of reflected and transmitted electrons
and positrons, we consider we have $M$ to be the total number of
the reflected and transmitted electrons (of both positive and
negative energies), and $N$ to be the total number of the
reflected and transmitted positrons (of both positive and negative
energies). Using (\ref{R,T-Dirac}) and the parameters in
(\ref{R',T'-Dirac}), we get
\begin{equation}\label{M,N-1}
R|_{E>0}+T'|_{E<0}=\Big(\frac{1+r}{1-r}\Big)^2+\frac{4kr}{(k-r)^2}=M,$$$$
R'|_{E<0}+T|_{E>0}=\Big(\frac{k+r}{k-r}\Big)^2+\frac{4r}{(1-r)^2}=N.
\end{equation}
The only task remaining is to prove $M=N$. The denominators for
the both relations in (\ref{M,N-1}) are the same, therefore we
shall switch to the numerators. We have
\begin{equation}\label{M,N-2}
\textmd{the numerator of
$M$}=k^2+r^3+2kr+k^2r^2+r^4+2kr^3+2k^2r+2r^3-12kr^2,$$$$
\textmd{the numerator of
$N$}=k^2+r^3+2kr+k^2r^2+r^4+2kr^3+2k^2r+2r^3-12kr^2.
\end{equation}
And this is what we were looking for; same values for the total
reflected and transmitted numbers, for both electrons and
positrons.

\section{Conclusion}
In this article we initially stated the Klein paradox, were the
anomalous and unexpected numbers of reflected electrons from a
potential barrier, has itself begun to appear as an obstacle
beyond physicists. Since Dirac himself had essentially omitted the
negative modes in his solutions, the most recognized explanation
for such paradox was the consideration of backwardly moving
electrons and transmitting positrons with positive energy. This
explanation and the similar ones, have been criticized and under
consideration for many years. Some physicists support theories
like them, however these are just legitimizations. In this
article, we are endeavoring to get rid of the paradox itself. We
believe that by maintaining the negative modes (or negative
energies), it would be possible to regain the total incidental
current and the exact number of electrons, without confronting
strange backward travelling electrons. Our progress was based on
the Krein quantization, concerning all four of Dirac's solutions.
What we are dealing with, is that we can validate this
possibility, that the negative states could be considered as viable energies.\\

Here we must note that, our approach to the quantum theory of
fields (namely Krein quantization), still exposes a mathematical
picture. Through this mathematical approach, some important
physical phenomena, like Casimir effect, have been retreated
\cite{137}. Briefly speaking, we try to remove the divergences,
caused by the QFT Green's function, by considering all four
solutions of Dirac field equation. This seems to be of some
physical applications and in this case, removing the Klein
paradox.\\

While the Krein quantization method, is a seemingly totally
mathematical one, some physicist are endeavoring to bring up its
probable physical implications (see \cite{138,139,140}). However,
despite these efforts, the proper physical picture of Krein
quantization is not still clear. Therefore, unless this method has
been used in order to explaining the QCD concepts and been
compared with the ghost field results, the appropriate physical
picture will not be clarified.\\

Nevertheless the negative energies in quantum physics have not
been related to real physical concepts, however the results in
this paper, may open doors for further considerations, and this is
what we were looking for.\\\\\\

\acknowledgements {\emph{We are grateful to the anonymous referee
for fruitful inquiries, which helped us improving the presentation
of the paper.}}

\end{document}